\documentclass[reprint,aip,apl,amsmath,amssymb]{revtex4-2}

\usepackage[dvipdfmx]{graphicx}
\usepackage{dcolumn}
\usepackage{bm}

\usepackage[utf8]{inputenc}
\usepackage[T1]{fontenc}
\usepackage{mathptmx}

\usepackage{ulem}
\usepackage{color}

\begin{document}

\preprint{AIP/123-QED}

\title{18.8 Gbps real-time quantum random number generator with a photonic integrated chip}

\author{Bing Bai}
 \affiliation{Hefei National Laboratory for Physical Sciences at the Microscale and Department of Modern Physics, University of Science and Technology of China, Hefei 230026, China}
 \affiliation{CAS Center for Excellence in Quantum Information and Quantum Physics, University of Science and Technology of China, Hefei 230026, China}
\author{Jianyao Huang}
 \affiliation{College of Information Science and Electronic Engineering, Zhejiang University, Hangzhou 310027, China}
\author{Guan-Ru Qiao}
 \affiliation{Hefei National Laboratory for Physical Sciences at the Microscale and Department of Modern Physics, University of Science and Technology of China, Hefei 230026, China}
 \affiliation{CAS Center for Excellence in Quantum Information and Quantum Physics, University of Science and Technology of China, Hefei 230026, China}
 \author{You-Qi Nie}
 \affiliation{Hefei National Laboratory for Physical Sciences at the Microscale and Department of Modern Physics, University of Science and Technology of China, Hefei 230026, China}
 \affiliation{CAS Center for Excellence in Quantum Information and Quantum Physics, University of Science and Technology of China, Hefei 230026, China}
 \author{Weijie Tang}
 \affiliation{College of Information Science and Electronic Engineering, Zhejiang University, Hangzhou 310027, China}
 \author{Tao Chu}
 \affiliation{College of Information Science and Electronic Engineering, Zhejiang University, Hangzhou 310027, China}
 \author{Jun Zhang}
 \email{zhangjun@ustc.edu.cn}
 \affiliation{Hefei National Laboratory for Physical Sciences at the Microscale and Department of Modern Physics, University of Science and Technology of China, Hefei 230026, China}
 \affiliation{CAS Center for Excellence in Quantum Information and Quantum Physics, University of Science and Technology of China, Hefei 230026, China}
 \author{Jian-Wei Pan}
 \affiliation{Hefei National Laboratory for Physical Sciences at the Microscale and Department of Modern Physics, University of Science and Technology of China, Hefei 230026, China}
 \affiliation{CAS Center for Excellence in Quantum Information and Quantum Physics, University of Science and Technology of China, Hefei 230026, China}

\date{\today}

\begin{abstract}
Quantum random number generators (QRNGs) can produce true random numbers.
Yet, the two most important QRNG parameters highly desired for practical applications, i.e., speed and size, have to be compromised during implementations. Here, we present the fastest and miniaturized QRNG with a record real-time output rate as high
as 18.8 Gbps by combining a photonic integrated chip and the technology of optimized randomness extraction.
We assemble the photonic integrated circuit designed for vacuum state QRNG implementation, InGaAs homodyne detector and high-bandwidth transimpedance amplifier into a single chip using hybrid packaging, which exhibits the excellent characteristics of integration and high-frequency response. With a sample rate of 2.5 GSa/s in a 10-bit analog-to-digital converter and subsequent paralleled postprocessing in a field programmable gate array, the QRNG outputs ultrafast random bitstreams via a fiber optic transceiver, whose real-time speed is validated in a personal computer.
\end{abstract}

\maketitle


Random numbers are widely required in diverse applications such as cryptography, simulation, lottery and scientific calculation.
Quantum random number generators (QRNGs) can produce true random numbers with characteristics of unpredictability, irreproducibility, and unbiasedness, which are guaranteed by the basic principle of quantum physics.
Over the last two decades, various QRNG schemes have been demonstrated \cite{herrerocollantes2017quantum,ma2016quantum}, such as the beam splitter scheme by measuring the path selection of single photons \cite{stefanov2000optical,jennewein2000fast},
the time measurement scheme by digitizing the arrival time of single photons \cite{dynes2008high,wayne2009photon,wahl2011ultrafast,nie2014practical}, the quantum phase fluctuation scheme by measuring phase fluctuations due to the spontaneous emission of laser
\cite{williams2010fast,qi2010high,xu2012ultrafast,nie2015generation,zhou2015randomness,zhang2016note,liu2016117,lei20208},
and the vacuum state scheme by measuring quantum noise fluctuations \cite{gabriel2010generator,symul2011real,shi2016random,zheng20196,zhou2019practical,haylock2019multiplexed}. Compared with the beam splitter scheme and the time measurement scheme,
the quantum phase fluctuation scheme and the vacuum state scheme can easily achieve random bit rates up to Gbps,
due to the fact that photodetectors instead of single-photon detectors are used in such schemes.
For instance, in the vacuum state scheme the quantum noise fluctuations are measured by a balanced homodyne detector.
The randomness comes from the quadrature measurement on the vacuum state, and the uncertainty relation guarantees the unpredictability of measurement outcome. Apart from the advantages of high speed, the vacuum state scheme requires relatively fewer components for QRNG implementation, which is favourable for integration.

For practical use, the most important parameters of QRNG are real-time output speed and module size.
On one hand, ultrafast QRNGs depend not only on the high bandwidth of entropy source \cite{nie2015generation,liu2016117}, but also on the high-speed acquisition and postprocessing electronics \cite{zhang2016note,zheng20196,lei20208}. Previously QRNG based on quantum phase fluctuation with the highest real-time rate of 8.4 Gbps has been demonstrated on a bulky experimental setup \cite{lei20208}.
On the other hand,
the emerging technologies of integrated quantum photonics have exhibited significant advantages in terms of size reduction for specific optical applications \cite{wang2020integrated}. Recently several integrated QRNG schemes have been demonstrated \cite{abellan2016quantum,raffaelli2018generation,raffaelli2018homodyne,roger2019real,huang2019integrated}.

Aiming for the photonic integration, different optical waveguide platforms can be used.
For instance, a QRNG based on lithium niobate platform has achieved a real-time rate of 3.08 Gbps by using multiplexed detectors \cite{haylock2019multiplexed}, and a quantum entropy source has been implemented on InP platform \cite{abellan2016quantum}.
Compared with the III-V platforms, the silicon-on-insulator (SOI) platform has better technical maturity and higher integration density. Recently QRNG implementations based on SOI platform have been reported with the digitization by oscilloscope \cite{raffaelli2018generation,raffaelli2018homodyne}. However, achieving both monolithic integration and high performance on one platform always exists technical difficulties in practice.
In the vacuum state scheme, the dark current, responsivity and bandwidth of photodetector can significantly affect the generation rate of QRNG.
In the SOI platform, Germanium (Ge) photodiode (PD) is often used for monolithic integration \cite{michel2010high}, however, the dark current of Ge PD is larger with several orders of magnitude than that of InGaAs PD. Considering the performance and technical complexity, hybrid photonic integration provides an effective approach to electro-optical integration applications, particularly to the implementations of integrated QRNGs.

In this letter, we present an integrated QRNG based on vacuum state fluctuations with a record real-time output rate as high as 18.8 Gbps. We assemble the photonic integrated circuit, InGaAs homodyne detector and high-bandwidth transimpedance amplifier (TIA) into a single chip with the technology of hybrid packaging. Further,
by combining the chip and high-speed random data acquisition and postprocessing, we implement an ultrafast and compact QRNG module.
The photonic integrated circuit is fabricated on an SOI platform, including a 2 $\times$ 2 multimode interference (MMI) coupler, two variable optical attenuators (VOA) and a grating coupler. An InGaAs homodyne detector is hybrid integrated into the photonic chip, and the TIA is connected with the electrodes of homodyne detector using wire bonding.
Due to photonic integration and excellent characteristics of InGaAs homodyne detector, the frequency response of quantum entropy source is significantly improved. With a sample rate of 2.5 GSa/s for random data acquisition and an optimized Toeplitz hashing algorithm for randomness extraction in a field programmable gate array (FPGA), the QRNG module can generate random bitstreams with ultrafast real-time speed. After transmitting the random numbers to a personal computer
via a fiber optic transceiver, all the final random bits can well pass the standard randomness tests.


\begin{figure}
\includegraphics[width=1\linewidth]{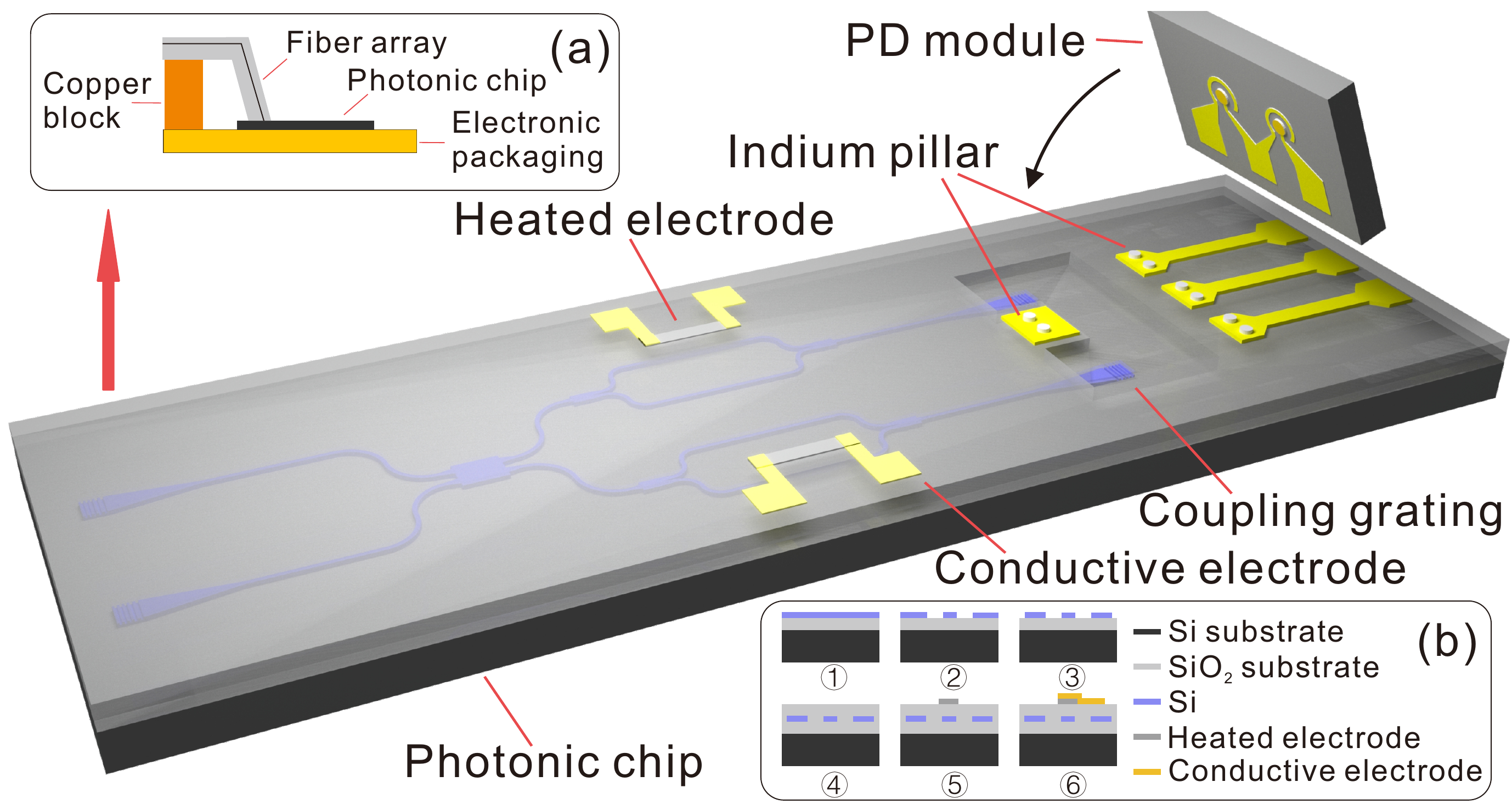}
\caption{\label{fig1} Structure and fabrication of the hybrid photonic chip. (a) Diagram of fiber array coupling.
    (b) Process flow chart for the fabrication of SOI chip.}
\end{figure}

Figure~\ref{fig1} depicts the structure layout of the hybrid photonic chip for the vacuum state scheme, with a size of 5 $\times$ 3 mm. The chip integrates one 2 $\times$ 2 MMI coupler, two VOAs, four coupling gratings and one homodyne PD module. The metal part inside the chip includes heated electrodes, thermally conductive electrodes and indium pillars. The incoming light is coupled into two coupling gratings via a single-mode fiber array, as shown in Fig.~\ref{fig1}(a). A copper block is used at the back of the fiber array and fixed with high-strength UV glue to support the fiber array, which guarantees the stability of the overall structure.
The MMI and VOAs are used to divide the incoming light into two paths and to independently adjust the optical attenuation in each path. Such design effectively ensures the length equality in both paths and thus the equal distribution at the two output ports. The homodyne InGaAs PD module is connected with the grating couplers by welding the electrodes on the photonic chip. The homodyne detection signal is drawn out through the common electrode of the PD module.

For the fabrication of SOI waveguide, the material structure and the process flow are illustrated in Fig.~\ref{fig1}(b). The SOI waveguide is fabricated on a silicon wafer with an insulating substrate (220 nm top silicon layer and 2 $\mu$m buried oxygen layer). After silicon electron beam exposure lithography and silicon etching process, silicon waveguides and thermal isolation grooves are prepared. The thermal isolation groove is used to isolate the heat generated by thermally tunable electrodes. Then, a 2 $\mu$m SiO$_{2}$ cladding layer is grown to protect the silicon waveguide. Further, we use lithography and SiO$_{2}$ etching technology to etch 1 $\mu$m thickness concave groove on the SiO$_{2}$ cladding layer for bonding the PD module. After such process, metal sputtering and metal stripping are used to sequentially grow the heated electrodes and conductive electrodes. Also, eight indium pillars are formed near the concave grooves for the circuit connection and the support of the PD module.

Considering the fact that InGaAs material has better performance superiority than silica-based material for high-speed photoelectric detection, an InGaAs PD module for homodyne detection is hybrid integrated. Inside the PD module, two co-directional InGaAs PDs are connected in series and the optical responsivity of both PDs reaches $\sim$ 0.95 A/W at 1310nm. With such design, the paired PDs exhibit almost the same optical-electrical characteristics, which is favourable for homodyne detection. The PD module is assembled on the surface of the SOI waveguide with flip-chip bonding, and the three electrodes of the PD module are correspondingly connected with the three metal electrodes on the photonic chip by indium pillars.

For optical coupling, a single-mode fiber array with a 90$^{\circ}$ bending structure is used and at the end of the fiber
a specific angle of 11.7$^{\circ}$ is designed for coupling the light from the output port of fiber array to the grating of the SOI waveguide. Such structure can both reduce the height of photonic chip package and realize the maximum coupling efficiency for the incoming light. Then, the fiber array and the photonic chip are fixed using ultraviolet glues with a refractive index of 1.450-1.455.
After the photonic chip is packaged, according to the photocurrent of PD and the intensity of incoming light the overall transmission loss is calculated to be $\sim$ 8.5 dB, including 0.2 dB insertion loss and 3 dB splitting ratio of the 2 $\times$ 2 MMI.


\begin{figure*}
\includegraphics[width=0.8\linewidth]{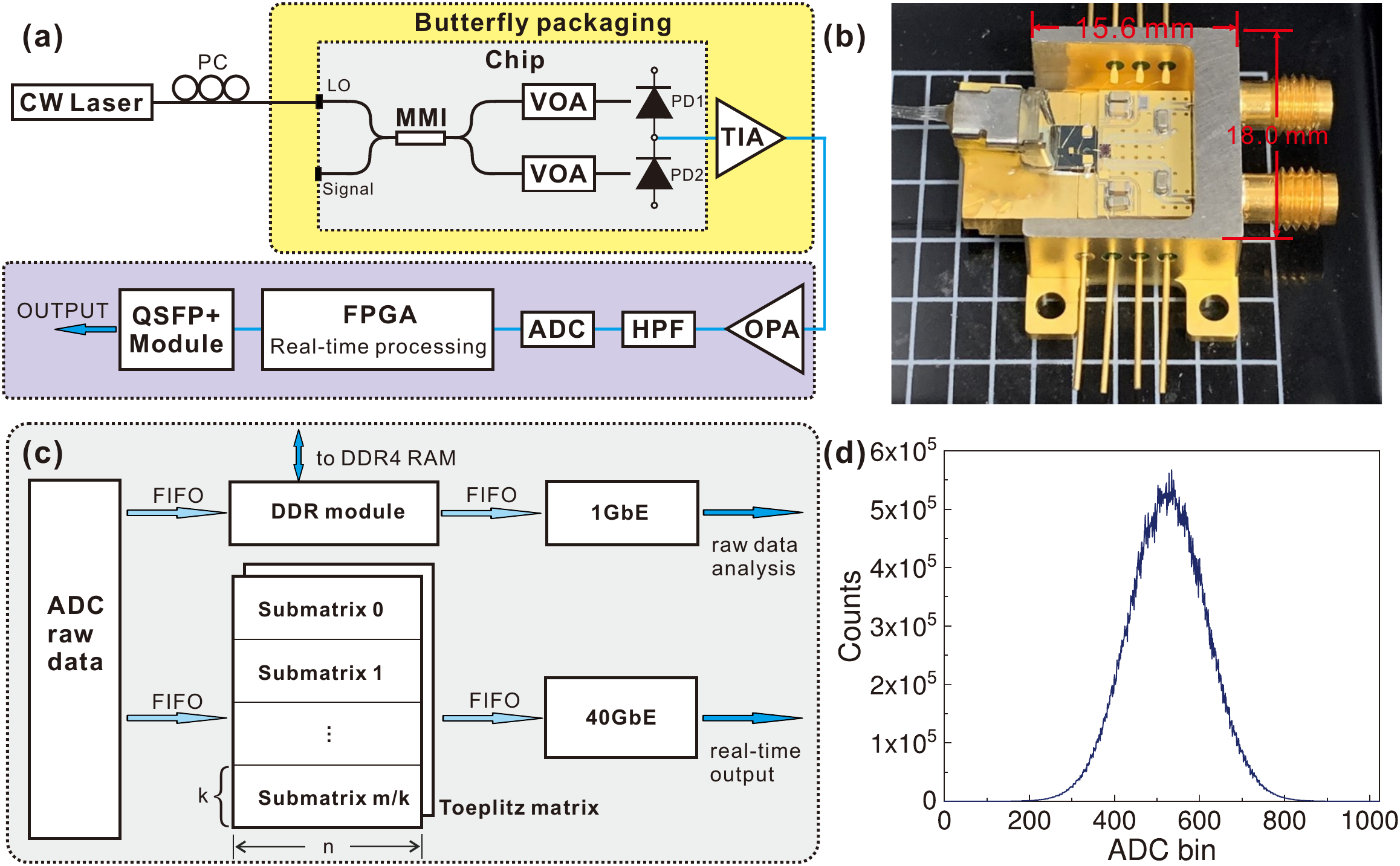}
\caption{\label{fig2}(a) Schematic diagram of the QRNG module. (b) Photo of the hybrid packaging. (c) Schematic diagram of random data processing including data acquisition, randomness extraction, and data transmission. (d) Raw data distribution sampled by the 10-bit ADC.}
\end{figure*}

Figure~\ref{fig2}(a) illustrates the experimental setup of QRNG module. A tunable continuous wavelength (CW) laser (Santec TSL-550-C) at 1310 nm is used as a local oscillator (LO), and connected with the fiber array via a polarization controller (PC) for optimizing the coupling efficiency. The other port of MMI is blocked as vacuum state preparation. A high-bandwidth TIA is connected with the photonic chip using wire bonding to minimize the parasitic parameters, so that the detected photocurrent difference from the PD module is amplified. Both the photonic chip and TIA
are mounted on a ceramic substrate and then assembled together within a butterfly package with a size of 15.6 mm $\times$ 18.0 mm, as shown in Fig.~\ref{fig2}(b). The butterfly package is further soldered on a printed circuit board with low-noise design.
In such a way, the entropy source of vacuum state can be operated in a pretty stable condition.

The output signal from the TIA is further amplified by an operational amplifier (OPA), and then filtered by a high-pass filter (HPF)
with a pass band from 1150 MHz to 5000 MHz
to match the measurement range of an analog-to-digital converter (ADC). The output signal from the HPF is digitized by the 10-bit ADC. The raw random data are sent into the FPGA for data analysis and real-time postprocessing based on Toeplitz matrix hashing. Both part of the raw data for analysis and the extracted random numbers are transmitted to a personal computer via an 1 GbE network interface and an 40 GbE quad small form-factor pluggable plus (QSFP+) module, respectively, as illustrated in Fig.~\ref{fig2}(c). Figure~\ref{fig2}(d) plots a typical Gaussian distribution of raw random data acquired by the ADC.

\begin{figure}
\includegraphics[width=1\linewidth]{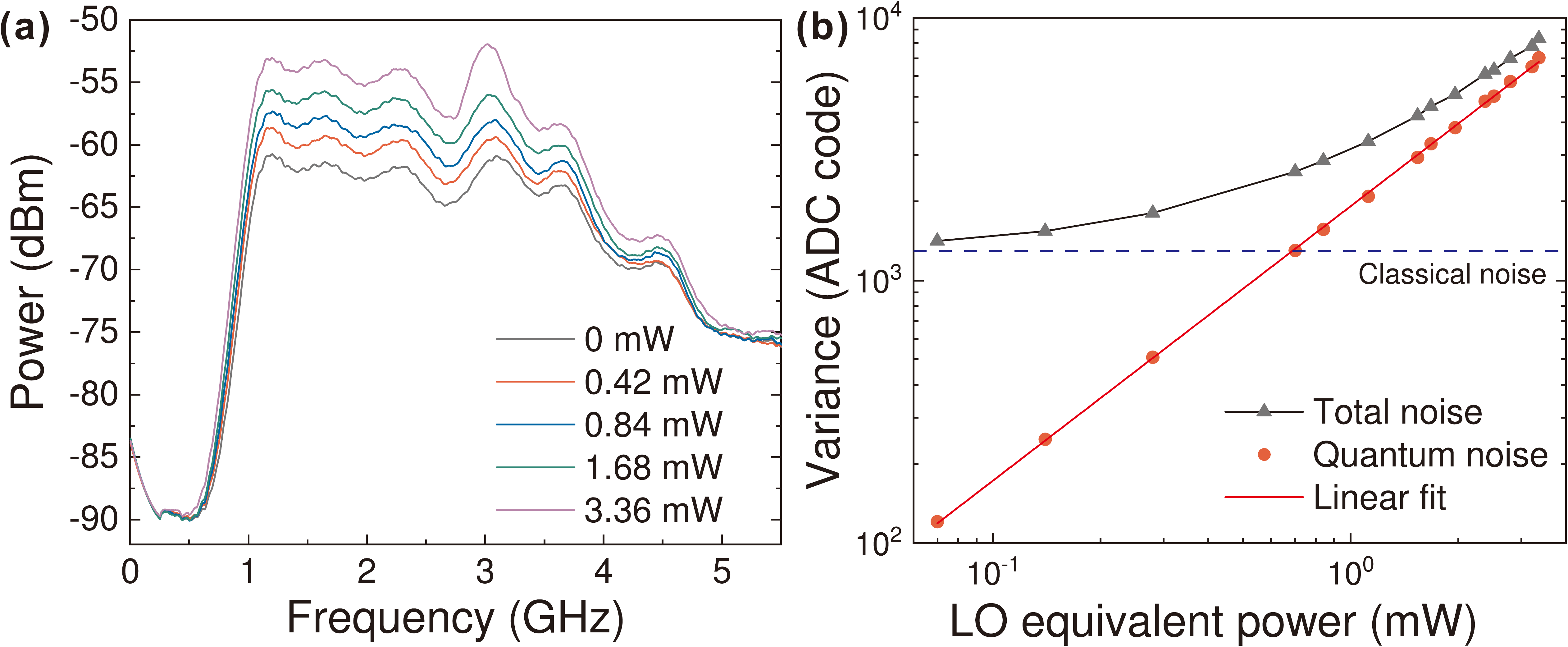}
\caption{\label{fig3}(a) Averaged power spectral density measured at the output of HPF with different settings of local oscillator equivalent power. (b) Variance of signal amplitude measured by the ADC as a function of LO equivalent power. The red line and the blue dot line represent the theoretical linear fit and the classical noise contribution, respectively.}
\end{figure}

In order to set an appropriate sampling rate for the ADC, we measure the averaged power spectral density through oscilloscope (Keysight MSOS804A) at the output of HPF with different LO equivalent power that is defined as the product of laser power at the LO port and the overall transmission loss of the photonic chip. The results are plotted in Fig.~\ref{fig3}(a), from which one can clearly observe the frequency response for the vacuum state fluctuations. On one hand, due to the HPF the lower frequency of output signals is cut to $\sim$ 1.2 GHz, and the most power of signals is distributed in the range from 1.2 GHz to $\sim$ 3.5 GHz, with constant signal-to-noise ratio (SNR) given a fixed LO equivalent power. When frequency is larger than $\sim$ 3.5 GHz, the SNR significantly reduces.
On the other hand, the SNR increases with an increase of LO equivalent power. The maximum LO equivalent power reaches 3.36 mW due to the power limit of CW laser.

Considering both the frequency response results in Fig.~\ref{fig3}(a) and the postprocessing capability limitation in FPGA, the sampling rate of the 10-bit ADC is set to 2.5 GSa/s.
The raw data of 25 Gbps digitalized by the ADC are parallelly feed into the FPGA. The FPGA first performs the deserialization for the incoming data. Then, a small portion of raw data are stored in double data rate (DDR) memory and further transmitted to the personal computer through the 1 GbE network for the min-entropy evaluation, while the remaining data are processed by Toeplitz matrix, as illustrated in Fig.~\ref{fig2}(c). Toeplitz matrix is a hashing extractor to distill raw random data. Given a binary Toeplitz matrix with a size of $m \times n$, $m$ final random bits are extracted by multiplying the matrix and $n$ raw bits.
Due to the computation capability limitation, the FPGA cannot directly perform the process for super large matrix. A concurrent pipeline algorithm is used instead to achieve the real-time extraction process by improving utilisation of FPGA resources \cite{zhang2016note}. The whole Toeplitz matrix is divided into $n/k$ submatrices, see Fig.~\ref{fig2}(c). Considering the resource consumption and raw data length in the FPGA, the values of $n$ and $k$ are set to be 1024 and 80, respectively. After the min-entropy evaluation, the factor $m$ can be determined. Finally, the extracted real-time random numbers are transmitted to the personal computer via the 40 GbE transceiver module.


The homodyne detector performs quadrature measurement on vacuum states. According to the uncertainty principle, the results of continuous measurement on orthogonal amplitude are unpredictable \cite{gabriel2010generator}. In reality, the result of homodyne detection includes other noise contributions such as the imperfection of the MMI splitting ratio, the difference of the PD responsivity and the intrinsic electronic noise \cite{masalov2017noise}.

Due to the imperfection during the SOI fabrication process, the split ratio of MMI cannot be 50:50 precisely, for which two thermally controlled Mach–Zehnder interferometers are used as VOAs in both paths to balance the splitting ratio of MMI.
The VOAs can be modelled as a part of beam splitter in homodyne detection \cite{huang2019integrated}. With the configuration of two VOAs, the optical path difference between two arms can be minimised, which is necessary to improve the SNR of homodyne detection.

Apart from the quantum shot noise in homedyne detection, the major contribution comes from the intrinsic electronic noise in PDs and the amplification circuit. The noise in PDs includes thermal noise and dark current noise, which both approximately follow Gaussian statistics.
Similarly, the noise in the amplification circuit also follows a Gaussian distribution at high frequency.

In order to quantify the randomness in the experiment, given that the quantum noise and the classical noise follow Gaussian distributions the measured total variance of signal amplitude can be described by
$\sigma_{total}^{2}=\sigma_{q}^{2}+\sigma_{c}^{2}$,
where $\sigma_{total}^{2}$ is the total variance of amplitude measured by the ADC, $\sigma_{q}^{2}$ is the quantum noise contribution, and $\sigma_{c}^{2}$ is the classical noise contribution. $\sigma_{c}^{2}$ is measured in the case without the LO input. By varying the LO input power, $\sigma_{total}^{2}$ changes. Figure~\ref{fig3}(b) shows the experimental results of variances in the cases of different LO equivalent powers. We note that ADC code is equivalent to the unit of signal amplitude, i.e., code 0 and code 1023 corresponding to the dynamic range of the ADC, and variance is calculated according to $10^7$ samples at each setting.

From Fig.~\ref{fig3}(b), one can observe that the variance of quantum shot noise increases linearly as the increase of equivalent power, which agrees well with the theoretical predictions in the QRNG scheme of vacuum state fluctuations.
When the LO equivalent power is set to 3.36 mW, the QRNG module is continuously operated to generate random numbers.
The randomness of the raw data is evaluated by min-entropy approach, which is calculated as $H_{\infty}(X)=-\log _{2} P_{\max }=7.71$ bits per sample according to the measured value of $\sigma_{q}$.
We note that in the current theoretical analysis, a non-adversarial model is considered. The extracted randomness includes both the intrinsic randomness originated from quantum shot noise and the nominal randomness from mixed states, where the latter might be controlled by classical or quantum side information \cite{zhou2018randomness,yuan2019quantum}. In principle, one can take account of the potential leakage of side information and quantify the amount of intrinsic randomness with condition min-entropy, which deserves future investigations.
For the real-time randomness extraction in the FPGA, the factor $m$ of Toeplitz matrix is set to 1360. Therefore, the randomness extraction ratio is $n/m=0.753$, and the final random bits are transmitted in real-time to the personal computer via the 40 GbE transceiver module. The generation rate is validated to be $\sim$ 18.8 Gbps.

For the postprocessing electronics, an FPGA evaluation board (Xilinx Virtex UltraScale VCU108) with two FMC connectors and one QSFP+ connector is used. The extracted random data is serialised and transmitted to a personal computer through the optical transceiver including four data channels with different wavelengths. The extracted data of $\sim$ 18.8 Gbps are transmitted via two data channels.
In the receiver, a wavelength-division multiplexer is used to separate optical signals and a two-channel 10 GbE network adapter (Intel X710) receives the incoming data. The real-time network transmission speed in each channel is continuously monitored in the task manager of Windows 10, with slight fluctuations around 9.4 Gbps in each channel. We also develop a software to further process the final random data, by which the data are cached into DDR4 memory first and then stored into a high-speed solid-state drive (Samsung 980 PRO) operated at PCIe 4.0 mode.

\begin{figure}
\includegraphics[width=1\linewidth]{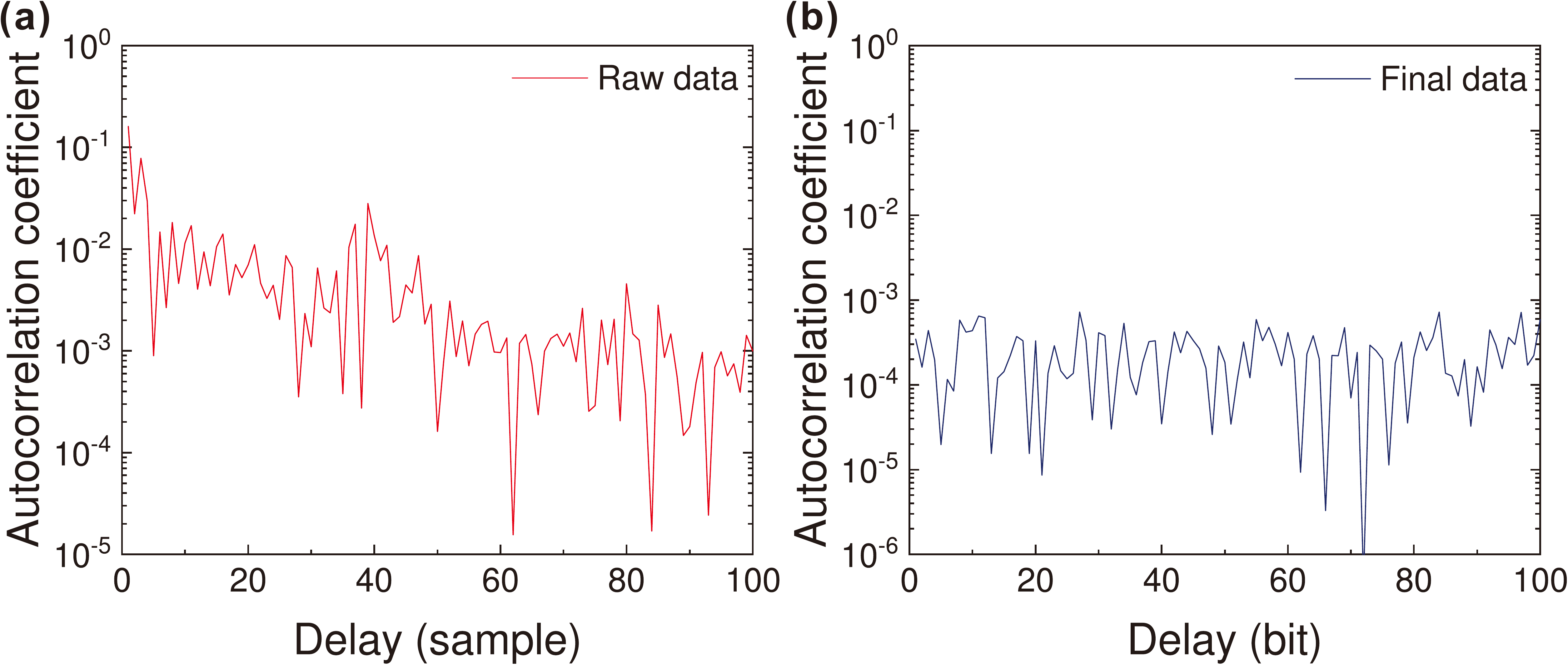}
\caption{\label{fig4}(a) Autocorrelation coefficient calculation of raw random data. The data size of each file is 10$^7$ samples. (b) Autocorrelations coefficient calculation of the final random data after extraction. The data size of each file is 10$^7$ bits.}
\end{figure}

To investigate the randomness of the final data, we perform an autocorrelation coefficient comparison between the raw random data and the extracted random data. Figure~\ref{fig4} exhibits the results in two cases, from which one can conclude that the slight autocorrelation in the range with small delay as shown in Fig.~\ref{fig4}(a) is significantly reduced after real-time extraction. Further, we perform the standard NIST statistical tests to verify the randomness of the final random data and the final random numbers well pass all the test items.

In conclusion, we have reported the fastest and miniaturized QRNG module with a record real-time output rate of 18.8 Gbps by combining a photonic integrated chip and optimized randomness extraction.
We have designed and fabricated a specific photonic integrated circuit for the QRNG scheme of vacuum state fluctuations, and further assembled the photonic integrated circuit, InGaAs homodyne detector and high-bandwidth TIA into a single chip using hybrid packaging, which greatly improves both the integration and high-frequency response of the QRNG module.
We have also implemented ultrafast randomness extraction in the FPGA with real-time postprocessing capability for a 10-bit ADC with
a sample rate of 2.5 GSa/s. After transmitting the final random bits to a personal computer via 40 GbE optical transceiver, the real-time random bit rate is validated.
Our work paves the way to implement an ultrafast and integrated QRNG module for practical applications.

\begin{acknowledgments}
We acknowledge technical supports from QuantumCTek Co., Ltd. and China Electronics Technology Group Corporation No.~44 Research Institute.
This work has been supported by the National Key R\&D Program of China under Grant No.~2017YFA0304004, the National Natural Science Foundation of China under Grant No.~11674307, the Chinese Academy of Sciences, and the Anhui Initiative in Quantum Information Technologies.
\end{acknowledgments}

\section*{DATA AVAILABILITY}
The data that support the findings of this study are available from the corresponding authors upon reasonable request.

\bibliography{ref}


\end{document}